\documentclass[letterpaper]{IEEEtran}
\usepackage{amsfonts}
\usepackage{mathrsfs}
\usepackage{amssymb,amsmath}
\usepackage[noblocks]{authblk}
\usepackage[compress]{cite}
\usepackage{bm}
\usepackage{algorithm}
\usepackage{algorithmic}
\usepackage{amsmath,amssymb,epsfig,graphics,subfigure}
\usepackage{theorem}
\usepackage{array,color}
\usepackage[compress]{cite}
\usepackage{algorithm}
\usepackage{verbatim}
\usepackage{mathrsfs}
\usepackage{amssymb}
\usepackage{graphicx}
\usepackage{float} 
\usepackage{amsmath} 
\usepackage[ textfont=normalfont, justification=centering]{caption}

\usepackage{theorem}
\usepackage{dsfont}

\allowdisplaybreaks

\usepackage{caption}
\theoremheaderfont{\normalfont\bfseries}
\usepackage{algorithmic}

\makeatletter
\newcommand{\removelatexerror}{\let\@latex@error\@gobble}
\makeatother

\makeatletter

\newcommand{\Rmnum}[1]{\expandafter\@slowromancap\romannumeral #1@}
\makeatother
\usepackage[numbers,sort&compress]{natbib}
\title{Movable Beyond-Diagonal Reconfigurable Intelligent Surfaces: Moving, Interconnecting, or Both? }
\author{
    \IEEEauthorblockN{Shuyue Xu, Matteo Nerini, {\em Member, IEEE}, and Bruno Clerckx, {\em Fellow, IEEE}}
    \thanks{
        S. Xu is with the 
		 School of Information and Electronics, Beijing Institute of Technology, Beijing 100081, China 
		 (e-mail: shuyuexu.ee@email.com).
    }
    \and
    \IEEEauthorblockN{ }
    		\thanks{
        M. Nerini and B. Clerckx are 
		 with the Department of Electrical and
Electronic Engineering, Imperial College London, London SW7 2AZ, U.K.
(e-mail:\{m.nerini20, b.clerckx\}@imperial.ac.uk).
    }
}

\begin{document}

\maketitle
\begin{abstract}
This letter proposes a  movable beyond-diagonal  reconfigurable intelligent surfaces (MA-BD-RIS) design, combining  inter-element connectivity  and  movability for  channel enhancement. We study a MA-BD-RIS assisted  multi-user multiple input single output system where beamforming, BD-RIS configuration, and elements positions are jointly optimized to maximize the sum-rate. An efficient  algorithm is developed, incorporating closed-form beamforming, a low-complexity partially proximal alternating direction method of multipliers for BD-RIS design, and successive convex approximation for  element placement. Simulations show that the high-movability structure yields superior
performance in  small-scale RIS and rich scattering scenarios, while the high-connectivity structure dominates in  large-scale RIS and massive transmit array configurations.
\end{abstract}
\section{Introduction}

Reconfigurable intelligent surface (RIS) has emerged as a pivotal technology for the 6G wireless networks, owing to its ability to reconfigure wireless channels in a cost- and energy-efficient manner \cite{QWUTCOM2021}. 
Both academia and industry have devoted considerable efforts to RIS research, spanning channel estimation, beamforming design, and prototyping  \cite{szhang2020jsac}. Owing to its ease of deployment and high flexibility in control, RIS can be utilized to bypass obstacles and dynamically manipulate the directivity of electromagnetic waves \cite{ZWangtwc2024}. 

The recently introduced movable antenna (MA) paradigm exploits the physical displacement of antenna to yield additional channel degrees of freedom, demonstrating significant potential for enhancing key system metrics like throughput and beamforming flexibility over fixed antenna (FA) systems \cite{Lzhu2025survey}. Inspired by this, the movability concept has been extended to RIS elements to eliminate phase distribution offset under discrete phase shifts \cite{Ghulett2024}. Furthermore, the MA-RIS structure has been proposed to dynamically adjust element positions for SNR enhancement and outage probability reduction \cite{Yzhang2024arxiv}.

 However, conventional RIS is limited by its  diagonal scattering matrix \cite{Hlijsac2023}. The beyond-diagonal RIS (BD-RIS) concept overcomes this by establishing inter-element connectivity \cite{SSHENTWC2022}, which provides enhanced degrees of freedom through a non-diagonal scattering matrix, significantly improving transmission quality and coverage \cite{Hliarxiv2025}.  BD-RIS performance improves with connectivity, and a performance-connectivity Pareto frontier trade-off is established in \cite{Nerini2023letter}.
 
  
 Stronger BD-RIS connectivity incurs substantial hardware complexity, often with diminishing performance returns. To address this, integrating BD-RIS with MAs offers a promising alternative, as the physical displacement of  elements can  compensate for reduced interconnections, simplifying hardware while maintaining channel flexibility. Given that the systematic potential of combining BD-RIS with MAs remains largely unexplored in existing studies, this paper utilizes
 group-connected MA-BD-RIS architecture  to investigate the performance trade-off between circuit connectivity and element movability. Within the architecture, 
 the lack of inter-group connectivity  in the group-connected BD-RIS provides
provides the feature of independent movability for each group.

The main contribution of this work lies in establishing a general modeling framework for MA-BD-RIS aided MU-MISO systems and proposing a low-complexity joint optimization algorithm by exploiting the coordinate optimization framework. Unlike conventional designs where MAs are deployed only at the transmitter or receiver side, our model accounts for BD-RIS element repositioning that simultaneously affects both transmit and receive channels, which introduces unique optimization challenges. Our comprehensive evaluations reveal the fundamental trade-off between circuit connectivity and element mobility, offering valuable design guidelines for future BD-RIS systems.
\vspace{-10pt}
\section{System Model and Problem Formulation}
\vspace{-15pt}
	\begin{figure}[htbp]
		\includegraphics[width =0.45\textwidth]{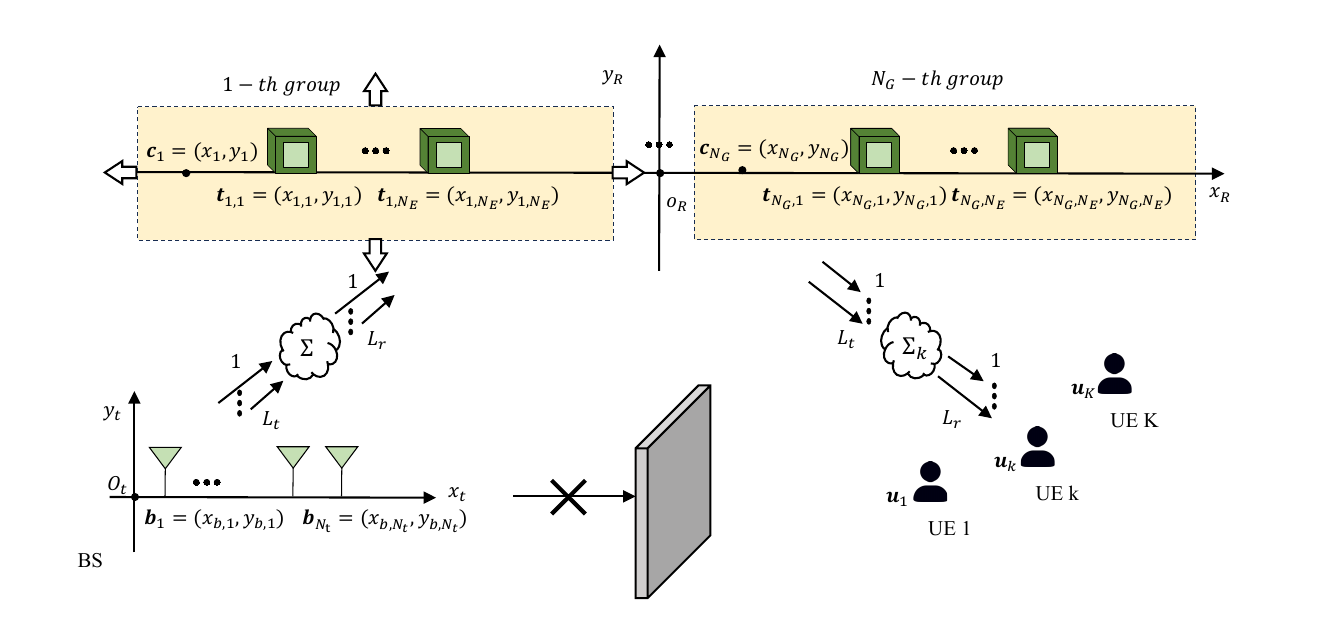}
		\caption{
			 Model of the MA-BD-RIS aided MU-MISO system.
		}
		\label{fig1}	
		\vspace{-10pt}
	\end{figure}
	
As illustrated in Fig.~1, we consider a MA-BD-RIS assisted MU-MISO downlink system, where a BS equipped with $N_t$ FAs communicates with $K$ fixed single-antenna UEs, aided by a  MA-BD-RIS with $M$ elements. The general group-connected RIS architecture is adopted to strike a balance between circuit complexity and performance, which includes the single-connected and fully-connected architectures as special cases \cite{SSHENTWC2022}. Specifically, the BD-RIS is divided into $N_G$ groups, each consisting of $N_E = \frac{M}{N_G}$ elements. The overall scattering matrix is modeled as a block-diagonal form ${\bm{\Theta}} = {\rm blkdiag} ({\bm{\Theta}}_{1},...,{\bm{\Theta}}_{N_G})\in\mathbb{C}^{M\times M}$, which satisfies ${\bm{\Theta}} = {\bm{\Theta}}^{\rm T}$ and ${\bm{\Theta}}^{\rm H}{\bm{\Theta}} = \bm{I}$. There is no inter-group connections between each group \cite{Hliarxiv2025}, thus, each group can be considered as an independent movable sub-panel, with  $N_E$ elements within the same group are rigidly fixed at half-wavelength spaced arrangement.

We employ local coordinate systems to define the spatial layout of the system. The  antenna positions for the BS are given by  $\bm{b}=[\bm{b}_{1},...,\bm{b}_{N_t}]\in\mathbb{C}^{2\times N_t}$ and the location of the $k$-th UE is  $\bm{u}_k\in\mathbb{C}^{2\times 1}$. Importantly, we assume that all antenna elements within the same BD-RIS group undergo collective movement as a single unit. Thus, the  position of elements in the $g$-th group  can be fully characterized  by a single reference point,  $\bm{c}_{g}=[x_{g},y_{g}]^{\rm T}$, which is set as the position of the first antenna element by default. Accordingly, the position of the $m$-th element in the $g$-th group is defined by $\bm{t}_{g,m}=\bm{c}_{g}+\bm{\delta}_{g,m}=[x_{g,m},y_{g,m}]^{\rm T}, m \in [1,N_E],g\in[1,N_G]$, where $\bm{\delta}_{g,m} = [\Delta x_{g,m},\Delta y_{g,m}]$ denotes the  fixed relative displacement of the $m$-th element with respect to  $\bm{c}_{g}$.    

 The transmitted signal at the BS can be  expressed as $\bm{x}  = \bm{W}\bm{s}= \sum_{k=1}^{K}\bm{w}_{k}s_k$, with the beamforming matrix $\bm{W} = [\bm{w}_{1}, \ldots, \bm{w}_{K}] \in \mathbb{C}^{N_t \times K}$ and the symbol vector $\bm{s} \in \mathbb{C}^{K \times 1}$ satisfying $\mathbb{E}[\bm{s}\bm{s}^{\rm H}] = \bm{I}_K$. Under the assumption of a blocked direct link, the received signal at the $k$-th UE is
	\begin{equation}
		{y}_k(\bm{c})=\bm{h}^{\rm H}_k(\bm{c})\bm{\Theta}\bm{H}(\bm{c})\bm{W}\bm{s}+{n}_k,
	\end{equation}
	where $\bm{h}_k^{\rm H}(\bm{c})$ and $\bm{H}(\bm{c})$ denote  the  channel from MA-BD-RIS to $k$-th UE, and from the BS to the MA-BD-RIS, respectively, with  $\bm{c}=\{\bm{c}_g|g\in[1,N_G]\}$  being the set of all reference points. The term $n_k \sim \mathcal{CN}(0,\sigma^2)$ represents  the additive white Gaussian noise (AWGN) at the $k$-th UE.
	Then, the SINR of $k$-th UE can be expressed as
	\begin{equation}
		\gamma_k = \frac{|\bm{h}^{\rm H}_k(\bm{c})\bm{\Theta}\bm{H}(\bm{c})\bm{w}_{k}|^2}{\sum_{i = 1, i\neq k}^{K}|\bm{h}^{\rm H}_k(\bm{c})\bm{\Theta}\bm{H}(\bm{c})\bm{w}_{i}|^2+ \sigma^2}.	\vspace{-8pt}
	\end{equation}
 	\subsection{Field-Response Based Channel Model}
	This work considers far-field narrowband communications with quasi-static block-fading.  Let $L_t$ and $L_r$ be the numbers of transmit and receive  paths. Based on the field-response model \cite{Lzhu2025survey},
  the channel from the BS to the $g$-th group MA-BD-RIS can be expressed as 
  \vspace{-5pt} 	
     \begin{equation}
    \bm{H}_g(\bm{c}_{g}) =  \bm{F}^{\rm H}(\bm{c}_{g})\bm{\Sigma}_{br}\bm{G}(\bm{b}), \label{channel2}
	\vspace{-5pt} 	
    \end{equation} 
	where the field-response matrix (FRM) in the receive region $\bm{F}(\bm{c}_{g}) = [\bm{f}(\bm{t}_{g,1}) ,...,\bm{f}(\bm{t}_{g,N_E})]\in \mathbb{C}^{L_{r}\times N_E}$ and transmit region  $\bm{G}(\bm{b}) = [\bm{g}(\bm{b}_{1}) ,...,\bm{g}(\bm{b}_{N_t}) ]\in \mathbb{C}^{L_t\times N_t}$ characterizes  the phase difference  caused by the displacement of the $g$-th group at $\bm{c}_g$ in the receive region and  the $N_t$ antennas in the transmit region.
  $\bm{\Sigma}_{br} \in \mathbb{C}^{L_r\times L_t}$ represents the  path response matrix (PRM). Let  $\phi_{p}^x$ and $\theta_{p}^x$, $x \in \{t,r\}$ denote  the azimuth and elevation angles of the  $p$-th transmit and receive path from the BS to MA-BD-RIS, corresponding to the angle of departure (AOD) when $x=t$ and the angle of arrival (AOA) when $x=r$, respectively. Then the field-response vector (FRV) of the $m$-th element in the transmit and receive region can be defined as
	\begin{align}
		\bm{g}(\bm{b}_{m})& = [e^{j\frac{2\pi}{\lambda}\rho^{t}_{1}(\bm{b}_{m})},...,e^{j\frac{2\pi}{\lambda}\rho^{t}_{L_{t}}(\bm{b}_{m})}]^{\rm T},\nonumber \\
		\bm{f}(\bm{t}_{g,m})& =[e^{j\frac{2\pi}{\lambda}\rho^{r}_{1}(\bm{t}_{g,m})},...,e^{j\frac{2\pi}{\lambda}\rho^{r}_{L_{r}}(\bm{t}_{g,m})}]^{\rm T},\vspace{-10pt} 
	\end{align}
where $\rho^{t}_{p}(\bm{b}_{m}) = x_{{\rm b}, m}sin\phi_{p}^t cos \theta_{p}^t +y_{{\rm b}, m} sin\theta_{p}^t$ and $\rho^{r}_{p}(\bm{t}_{g,m}) = x_{g,m}sin\phi_{p}^r cos \theta_{p}^r +y_{g,m} sin\theta_{p}^r$. 
 Thus, the  channel between the  BS and the MA-BD-RIS can  be expressed as $\bm{H} = [\bm{H}_1(\bm{c}_{1}) ;...;\bm{H}_{N_G}(\bm{c}_{N_G})]$. 

The overall channel between the MA-BD-RIS and $k$-th UE can be expressed as $ \bm{h}_{k}(\bm{c}) = [{\bm{h}}_{k,1}(\bm{c}_{1})^{\rm T},...,\bm{h}_{k,N_G}(\bm{c}_{N_G})^{\rm T}]^{\rm T}$, and the channel with respect to the $g$-th group is   
\vspace{-3pt} 
	\begin{equation}
		 {\bm{h}}_{k,g}(\bm{c}_{g}) = \bm{G}_{k}(\bm{c}_{g})^{\rm H}\bm{\Sigma}_k^{\rm H} \bm{1}, \label{channel1}\vspace{-4pt} 
	\end{equation}
	where the transmit FRM $\bm{G}_{k}(\bm{c}_{g}) = [\bm{g}_{k}(\bm{t}_{g,1})  ,...,\bm{g}_{k}(\bm{t}_{g,N_E}) ]\in \mathbb{C}^{L_t\times N_E}$, and $\bm{1}\in \mathbb{C}^{L_r \times 1}$ is the all one element matrix.	\vspace{-15pt}
\subsection{Problem formulation} 
To gain more intuitive insight of the considered system, we assume the CSI is perfectly available at both transmitter and receiver side. In this paper, we aim to maximize the system sum rate  by cooperatively optimizing the beamformer matrix $\bm{W}$, scattering matrix $\bm{\Theta}$ and the reference point of each group $\bm{c}$. The problem can be formulated as
	\vspace{-8pt}
\begin{align}
    \max \limits_{\bm{c},\bm{\Theta},\bm{W}} \  &\sum_{k=1}^{K}{\log}_2 \left(1+ \gamma_k \right), \label{obj_1}\\
      s.t.\  & {\rm Tr}(\bm{W}^{\rm H}\bm{W}) \leq P, \tag{\ref{obj_1}{a}} \label{obj_1a}\\
      & \bm{c}_g \in \mathcal{C}_R,  g = 1,...,N_G,
	  \tag{\ref{obj_1}{b}} \label{obj_1b}
	   \\  
	& \Vert \bm{c}_g-\bm{c}_{g'} \Vert_2 \geq D,  \ g,g' = 1,...,N_G, g\neq g' ,
	  \tag{\ref{obj_1}{c}} \label{obj_1c}
	  \\
	  & {\bm{\Theta}} \!\!=\! \!{\rm blkdiag} ({\bm{\Theta}}_{1},...,{\bm{\Theta}}_{N_G}), \bm{\Theta}^{\rm H}\bm{\Theta} \!\!=\!\! \bm{I}, \bm{\Theta}^{\rm T} \!\!=\!\! \bm{\Theta},
	    \tag{\ref{obj_1}{d}} \label{obj_1d}
			\vspace{-10pt} 
\end{align}   
where (\ref{obj_1a}) restricts the maximal transmit power $P$,  (\ref{obj_1b}) confines the movement of each group within the predefined region $\mathcal{C}_R$, (\ref{obj_1c}) sets the minimum distance between different groups to $D$, (\ref{obj_1d}) reveals the nature of the group-connected BD-RIS. In order to be applicable to any given BD-RIS architecture, inspired by the work in \cite{zheyuarxiv2024}, we relate the scattering matrix $\bm{\Theta}$ with the admittance matrix $\bm{Y}\in \mathbb{C}^{M \times M}$ as
	\vspace{-3pt}
\begin{equation}
	\bm{\Theta} = (\bm{I}+Z_0 \bm{Y})^{-1}(\bm{I}-Z_0 \bm{Y}),	\vspace{-8pt}
\end{equation}
where the reference impedance $Z_0$ is set as $50 \Omega$. 
The admittance matrix $\bm{Y}$ should be set as a purely imaginary matrix, i.e., $\bm{Y}=j\bm{B}$, in order to maximize the reflected power. Thus, the original problem (\ref{obj_1}) can be transformed as
\begin{align}
	\vspace{-3pt} 
    \max \limits_{\bm{c},\bm{\Theta},\bm{B},\bm{W}} \  &\sum_{k=1}^{K}{\log}_2 \left(1+ \gamma_k \right), \label{obj_2}\\
      s.t.\  
	   & \bm{\Theta} = (\bm{I}+jZ_0\bm{B})^{-1}(\bm{I}-jZ_0\bm{B}),\tag{\ref{obj_2}{a}} \label{obj_2a}
	   \\
	   & {\bm{B}} = {\rm blkdiag} ({\bm{B}}_{1},...,{\bm{B}}_{N_G}),  \bm{B} = \bm{B}^{\rm T},
	  \tag{\ref{obj_2}{b}} \label{obj_2b}
	  \\
	  & (\ref{obj_1a})-(\ref{obj_1c}).
	  \tag{\ref{obj_2}{c}} \label{obj_2c}
\end{align} 
\vspace{-5pt}
\vspace{-20pt}
\section{ Algorithm Design}
In this section, we first employ the fractional programming (FP) technique \cite[Section IV]{ShenTSP2018} to transform the problem into a more trackable form and then  develop an alternative  algorithm to solve the problem.   
By introducing  auxiliary variables $\rho_k$ and $\psi_k$, the original problem (\ref{obj_2}) can be reexpressed as
	\vspace{-5pt}
\begin{align}
    &\max_{\substack{\rho_k,{\psi}_k,\bm{c},\\\bm{\Theta},\bm{B},\bm{W}}} 
     \sum_{k=1}^{K} [ \log_2(1\!+\!\rho_k)\!- \!\rho_k + 2\sqrt{1\!+\!\rho_k} \Re\left\{ \bm{h}_{k}^{\rm H}\bm{\Theta}\bm{H}\bm{w}_{k}{\psi}_k^* \right\}  \nonumber \\
    &  \!-\! |{\psi}_k|^{2} ( \sum_{i=1}^{K}\!|\bm{h}_{k}^{\rm H}\bm{\Theta}\bm{H}\bm{w}_{i}|^2 \!+ \!\sigma^2 \!) ], 
    {\rm{s.t.}}\  \rho_k \geq 0, (\ref{obj_2a})\!-\!(\ref{obj_2c}).
\end{align}
	In the following, we  alternately optimize each variable in $\{\rho_k,\psi_k,\bm{W},\bm{\Theta},\bm{B},\bm{c}\}$ while keeping the other fixed.
	\vspace{-11pt}
\subsection{Optimization of $\rho_k$ and ${\psi}_k$}
Given the other variables, the problem with respect to $\rho_k$ and ${\psi}_k$ is convex \cite{ShenTSP2018}. Exploiting the first-order optimal condition, the optimal solution can be derived as 
$
\rho_k^{opt} = \frac{|\bm{h}_{k}^{\rm H}\bm{\Theta}\bm{H}\bm{w}_{k}|^2}{\sum_{i=1, i\neq k}^{K}|\bm{h}_{k}^{\rm H}\bm{\Theta}\bm{H}\bm{w}_{i}|^2+\sigma^2}$ and $ \psi_k^{opt} = \frac{\sqrt{1+\rho_k}\bm{h}_{k}^{\rm H}\bm{\Theta}\bm{H}\bm{w}_{k}}{\sum_{i=1}^{K}|\bm{h}_{k}^{\rm H}\bm{\Theta}\bm{H}\bm{w}_{i}|^2+\sigma^2}$. 
	\vspace{-5pt}
\subsection{Optimization of $\bm{W}$} The problem with respect to  $\bm{W}$ can be written as
		\begin{equation}
		{\min_{\bm{W}}} \ 
		\sum_{k=1}^{K}\bm{w}_{k}^{\rm H} \bm{Q} \bm{w}_{k}-2\Re\{\bm{w}_{k}^{\rm H}{\bm{q}_k}\} 
		\ {\rm{s.t.}}\ \sum_{k=1}^{K}\bm{w}_{k}^{\rm H}\bm{w}_{k}\leq P,
	\end{equation}
	where $\bm{Q} = \sum_{i=1}^{K}|{\psi}_i|^{2}\bm{h}_{i}^{\rm H}\bm{\Theta}\bm{H}\bm{H}^{\rm H}\bm{\Theta}^{\rm H}\bm{h}_{i}$ is a positive semi-definite matrix and $\bm{q}_k =\sqrt{1+\rho_k}{\psi}_k\bm{H}^{\rm H}\bm{\Theta}^{\rm H}\bm{h}_{k}$. The problem is a quadratically constrained quadratic programming (QCQP). Based on the first-order optimality condition, the optimal solution  can be derived as 
	$\bm{w}_{k}^{\rm opt} = (\bm{Q}+\lambda \bm{I})^{-1}{\bm q}_k$,
	where $\lambda \geq 0$ is the Lagrange multiplier associated with the power constraint and can be obtained via a one-dimensional bisection search \cite{zheyuarxiv2024}. 
		\vspace{-12pt}
\subsection{Optimization of $\bm{\Theta}$ and $\bm{B}$}
The problem with respect to the scattering matrix $\bm{\Theta}$ and admittance matrix $\bm{B}$ is hard to solve due to the highly coupled variables and the nonconvex constraint in (\ref{obj_2a}). 
Inspired by  \cite{zheyuarxiv2024}, we introduce the auxiliary variables $\bm{u}_k=\bm{\Theta}^{\rm H} \bm{h}_k \in \mathbb{C}^M$ to eliminate the matrix inverse and to reduce the dimension of the optimized variable and constraints. Then, the problem can be expressed as 
	\begin{align}  \label{opt_b}
		{\max_{\bm{B},\bm{U}}} 
		& \tilde{R}(\!\bm{U}\!)\!\!=\!\!\!\sum_{k=1}^{K}\!  
		2\!\sqrt{\!1\!+\!\rho_k}\!
		\Re\{\!\bm{u}_k^{\rm H}\!\bm{H}\bm{w}_{k}{\psi}_k^*\!\}
		 \!\!-\!|{\psi}_k|^{2}\!(\sum_{i=1}^{K}\!|\!\bm{u}_k^{\rm H}\bm{H}\bm{w}_{i}|^2\!)
		, \nonumber \\
		{\rm{s.t.}}\ & \left(\bm{I}-j Z_0 \bm{B}\right) \bm{U}=\left(\bm{I}+j Z_0 \bm{B}\right) \bm{H}_{\rm U},(\ref{obj_2b}),
	\end{align}
where  $\bm{U}\!=\!\left[\bm{u}_1, \ldots, \bm{u}_K\right] \!\in\! \mathbb{C}^{M \times K} $ and $ \bm{H}_{\rm U}\!=\!\left[\bm{h}_1, \ldots, \bm{h}_K\right] \in \mathbb{C}^{M \times K}$. The problem is hard to solve due to the bi-linear constraint between the variable $\bm{U}$ and $\bm{B}$. Thus, we resort to the ADMM framework to handle the structure \cite{Stephenadmm}. The Lagrange dual function with respect to (\ref{opt_b}) can be written as
\begin{align}
& \mathcal{L}_\rho\!( \!\bm{B},\! \bm{U}\!,\! \bm{\Lambda})
\!\!=\!\!  \tilde{R}(\!\bm{U}\!)\!\!-\!
\!\Re\!\left\{\!{\rm Tr}\{\!\bm{\Lambda}^{\!\rm H}\!\!\left[\left(\bm{I}\!\!-\!j Z_0 \bm{B}\right) \!\bm{U}\!\!-\!\left(\bm{I}\!\!+\!\!j Z_0 \bm{B}\right) \!\!\bm{H}_{\rm U}\!\right]\!\}\!\!\right\}
\nonumber \\
& \!-\!\frac{\rho}{2}\!\left\|\left(\bm{I}\!-\!j Z_0 \bm{B}\right)\! \bm{U}\!\!-\!\left(\bm{I}\!+\!j Z_0 \bm{B}\right)\! \bm{H}_{\rm U}\!\right\|_F^2,
\end{align}	
where $\bm{\Lambda}$ is the Lagrange multiplier and $\rho$ is
 the penalty parameter. The partially proximal ADMM (pp-ADMM)  algorithm is given as follows
\begin{subequations}
	\vspace{-8pt} 
\begin{align}
& \bm{B}^{t+1} \!\in \!\underset{\bm{B}=\bm{B}^T,\bm{B} \in \mathcal{B}}{\arg \max } \mathcal{L}_\rho\left( \bm{B}, \bm{U}^t, \bm{\Lambda}^t\right)-\frac{\xi}{2}\left\|\bm{B}-\bm{B}^t\right\|_F^2,\label{obj_3a} \\
& \bm{U}^{t+1} \!\in \!\arg \max \mathcal{L}_\rho\!\left(\bm{B}^{t+1}, \bm{U}, \bm{\Lambda}^t\right),\label{obj_3b}\\
& \Lambda^{t+1}\!\!=\!\!\Lambda^t\!\!+\!\!\rho\!\left(\!\left(\bm{I}\!\!-\!\!j Z_0 \bm{B}^{t+1}\right) \!\!\bm{U}^{t\!+\!1}\!\!\!-\!\!\left(\bm{I}\!+\!j Z_0 \bm{B}^{t\!+\!1}\right) \!\!\bm{H}_{\rm U}\!\right)\label{obj_3c},
\end{align}
\vspace{-3pt}
\end{subequations}
where $\xi$ is  the corresponding proximal parameter, exploited to enhance stability and ensure the convergence of the  algorithm. 

     $1)$Update of $\bm{B}$: Considering the real-valued feature of $\bm{B}$, the B-subproblem can be expressed as  \vspace{-3pt} 
\begin{equation}
\min _{\bm{B}} \frac{\rho}{2}\|\bm{B} \bm{M}-\bm{\Gamma}\|_F^2+\frac{\xi}{2}\left\|\bm{B}-\bm{B}^t\right\|_F^2, \ {\rm s.t}. \ (\ref{obj_2b}), \label{obj4} 
\end{equation}
where $\bm{M}\!=\!\left[\Re\left(j Z_0 \bm{U}^t\!\!+\!\!j Z_0 \bm{H}_{\rm U}\right),\! \mathcal{I}\!\left(j Z_0 \bm{U}^t\!\!+\!\!j Z_0 \bm{H}_{\rm U}\right)\right] $ and  $\bm{\Gamma}=[\Re\left(\bm{U}^t-\bm{H}_{\rm U}+\frac{\bm{\Lambda}^t}{\rho}\right)$, $\mathcal{I}\left(\bm{U}^t-\bm{H}_{\rm U}+\frac{\bm{\Lambda}^t}{\rho}\right)]$. To further reduce the dimension of the variable, we exploit the block diagonal and real symmetric natural of the $\bm{B}$ to transform the problem (\ref{obj4}) as
\begin{equation}
\min _{\bm{x}}\frac{\rho}{2}\|\bm{A x}-\bm{b}\|_2^2+\frac{\xi}{2}\left\|\bm{x}-\bm{x}^t\right\|_2^2,\label{obj5}
\end{equation}
where $\bm{x}$ collects all non-zero elements in the upper triangular part of $\bm{B}$. The problem (\ref{obj5}) is now a unconstrained quadratic problem, and the optimal closed form is 
$
\bm{x}^{t+1}=\left( \bm{A}^T \bm{A}+\frac{\xi }{\rho}\bm{I}\right)^{-1}\left( \bm{A}^T \bm{b}+\frac{\xi }{\rho} \bm{x}^t\right).
$
The mapping relation between  $\bm{x}$ and $\bm{B}$, $\bm{A}$ and $\bm{M}$ is detailed in \cite{zheyuarxiv2024} and is not repeated here for brevity.

$2)$Update of $\bm{U}$: The problem with each $\bm{u}_k$ is separable and an unconstrained convex quadratic programming. Thus, the optimal closed solution is 
\begin{align}
\!\!\!\!&\bm{u}_k^{t+1}\!\!= \!\! \left(\left|\psi_k\right|^2 \!\bm{H}\bm{W}\bm{W}^{\rm H} \bm{H}^{\rm H} \! \!+\!\!\frac{\rho}{2}\left(\bm{I}+Z_0^2\left(\bm{B}^{t+1}\right)^2\right)\right)^{-1}
\nonumber \\
\!\!\!\!\!& 
\left(\! \!\!\sqrt{1\!+\!\rho_k}\!{\psi}_k^{*} \!\bm{H}\!\bm{w}_k\!\!+\!\!\frac{\rho\!\left(\bm{I}\!\!+\!\!j Z_0 \bm{B}^{t\!+\!1}\right)^{\!\!2}\!\! \bm{h}_k\!\!-\!\!\left(\bm{I}\!\!+\!\!j Z_0 \bm{B}^{t\!+\!1}\right)\!\!\bm{\lambda}_k^t}{2}\!\!\!\right)\!,\!\!\!
\end{align}
where $\bm{\lambda}_k$ denotes the $k$-th column of $\bm{\Lambda}$.
\vspace{-10pt} 	
\subsection{ Optimization of $\bm{c}$}	The problem with respect to $\bm{c}$ can be expressed as 
	\begin{align}\label{opt_t}
		{\max_{\bm{c}}} \ 
		& \sum_{k=1}^{K}
		2\sqrt{1+\rho_k}
		\Re\{\bm{h}_{k}^{\rm H}(\bm{c})\bm{\Theta}\bm{H}(\bm{c})\bm{w}_{k}{\psi}_k^*\}
		 \nonumber \\ &-\!\!|{\psi}_k|^{2}(\sum_{i=1}^{K}|\bm{h}_{k}^{\rm H}(\bm{c})\bm{\Theta}\bm{H}(\bm{c})\bm{w}_{i}|^2)
		, 
		{\rm{s.t.}}   (\ref{obj_1b})\!\!- \!\!(\ref{obj_1c}).
	\end{align}
	Due to highly coupled variable multiplications in the objective and constraint (\ref{obj_1c}), we optimize $\bm{c}_g$ with other groups $\bm{c}_{g'}$ ($g'\neq g$) fixed alternately until convergence.
	Substituting (\ref{channel2}) and (\ref{channel1})  into the objective function  and ignoring the constant term,  we have
\begin{align} \label{u_g}
\!\!\!\!\mu \! \left(\!\bm{c}_g\! \right) \!\!=\!\!\!&\sum_{k=1}^{K}
(\!-\!\!\sum_{k'=1}^{K}\!\!|\!{\bm{f}}^{\rm H}\!(\!\bm{c}_{g}\!){\bm{C}}_{g,k,k'}{\bm{g}_{k}}(\bm{c}_{g})|^2 )\!\!+\!\!2\Re\!\{ \!
		{\bm{f}}^{\rm H}\!(\!\bm{c}_{g}\!){\bm{E}}_{g,k}{\bm{g}_{k}}(\!\bm{c}_{g}\!)\! \}
		\nonumber \\
\overset{(a)}{=}& \!\!\!-\!\!\!\!\!\!\!\!\!\!\!\!\!\sum_{\substack{1 \le i,j,p,q \le L \\ 1 \le k,k' \le K}} \!\!\!\!\!\!\!\!\!\!
|c_{ij}^{gkk'}\!\!|
|c_{qp}^{gkk'}\!|cos(\kappa_{ijpq}^{kk'}(\!\bm{c}_{g}\!)\!)
\!\!+\!\!2 \!\!\!\!\!\! \!\sum_{\substack{1 \le i,j \le L \\ 1 \le k \le K}}\!\!\!\!\!\!|e^{gk}_{ij}|cos(\kappa^{k}_{ij}(\bm{c}_{g}\!)\!),\!\!\!\!
\end{align}
 where $c_{ij}^{gkk'} \triangleq [\bm{C}_{g,k,k'}]_{i,j}$, $e_{ij}^{gk} \triangleq [\bm{E}_{g,k}]_{i,j}$ with matrices $\bm{C}_{g,k,k'}$ and  $\bm{E}_{g,k}$ given in Appendix A.
 Equation  $(a)$ is established by bringing FRV ${\bm{f}}(\bm{c}_{g})$ and $\bm{g}_{k}(\bm{c}_{g})$ back.  	The variable is intricately nested as $\kappa_{ijpq}^{kk'}(\bm{c}_g) = \frac{2\pi}{\lambda}[-\rho^{r}_{i}(\bm{c}_{g}) +\rho^{t}_{k,j}(\bm{c}_{g})-\rho^{t}_{k,p}(\bm{c}_{g})+\rho^{r}_{q}(\bm{c}_{g})]+\angle c_{ij}^{gkk'} -\angle c_{qp}^{gkk'} $ and $\kappa^{k}_{ij}(\bm{c}_g) = \frac{2\pi}{\lambda}[-\rho^{r}_{i}(\bm{c}_{g}) +\rho^{t}_{k,j}(\bm{c}_{g})]+\angle e_{ij}^{gk}$. However, we can utilize the second-order Taylor expansion to construct the surrogate function of $\mu\left(\bm{c}_g \right)$. By introducing  ${\delta_{g}} \bm{I}_2 \succeq$ $\nabla^2  \mu(\bm{c}_g)  $, we have \vspace{-10pt} 	
	\begin{align}\label{loc_obj}\!  
	\!\mu\left(\bm{c}_g\right) \!\geq & \mu\left(\bm{c}_g^{(i)}\right)\!\!+\!\!\nabla \!\mu\!\!\left(\bm{c}_g^{(i)}\right)^{\!\!T}\!\!\!\left(\bm{c}_g\!\!-\!\!\bm{c}_g^{(i)}\!\right)  \!\!-\!\!\frac{\delta_{g}}{2}\!\!\left(\bm{c}_g\!\!-\!\bm{c}_g^{(i)}\!\right)^{\!\!T}\!\!\!\left(\bm{c}_g\!\!-\!\bm{c}_g^{(i)}\! \right)  \nonumber \\
	= & \!\!-\!\!\frac{\delta_{g}}{2} \bm{c}_g^T \bm{c}_g+\left(\!\nabla\!\mu\left(\bm{c}_g^{(i)}\right)\!\!+\!\!\delta_{g} \bm{c}_g^{(i)}\right)^{T\!\!} \!\!\bm{c}_g 
	\!\!+\!\text{constant}.\vspace{-10pt} 	
	\end{align}
	To improve readability, we define $\kappa_{x}^{y}(\bm{c}_{g}) = x_g*\beta_x^{y} + y_g*\gamma_x^{y}+ \eta_x$, where ${x},y$ denote the  corresponding index set. Then,  
	 the gradient vector $\nabla  \mu(\bm{c}_g) \!
		= \left[\frac{\partial \mu(\bm{c}_g) }{\partial x_g}, \frac{\partial \mu(\bm{c}_g) } {\partial y_g}\right]^T$ is expressed  in (\ref{eq_long}) presented on the following page.
		\begin{figure*}[!t] 
\normalsize 
\vspace*{-10pt}
\begin{equation}
\frac{-2\pi}{\lambda}\![-\!\!\!\!\!\!\!\!\!\!\!\!\sum_{\substack{1 \le i,j,p,q \le L \\ 1 \le k,k' \le K}}\!\!\!\!\!\!\!\!\!\!|c_{ij}^{gkk'}\!||c_{qp}^{gkk'}\!|\beta_{ijpq}^{kk'}\!\sin(\kappa_{ijpq}^{kk'}\!(\bm{c}_g\!)\!)
		\!+\!2\! \!\!\! \!\!\!\sum_{\substack{1 \le i,j \le L \\ 1 \le k\le K}}\!\!\!\!\!\! |e_{ij}^{gk} \!|\beta_{ij}^{k}\!\sin\!(\!\kappa_{ij}^{k}\!(\!\bm{c}_g\!)\!)
		,\!-\!\!\!\!\!\!\!\!\!\!\!\!\sum_{\substack{1 \le i,j,p,q \le L \\ 1 \le k,k' \le K}}\!\!\!\!\!\!\!\!\!\!|c_{ij}^{gkk'}\!\!||c_{qp}^{gkk'}\!|\gamma_{ijpq}^{kk'}\!\sin(\!\kappa_{ijpq}^{kk'}(\!\bm{c}_g\!)\!)
		\!+\!\!2\!\!\!\!\!\! \sum_{\substack{1 \le i,j \le L \\ 1 \le k\le K}}\!\!\!\!\!\!|e_{ij}^{gk} |\gamma_{ij}^{k}\sin(\kappa_{ij}^{k}(\bm{c}_g))]^T
\label{eq_long}
\end{equation}
\hrulefill 
\vspace*{-15pt} 
\end{figure*}
	Inspired by the inequality $\left\|\nabla^2 \mu(\bm{c}_g) \right\|_2 \leq\left\|\nabla^2 \mu(\bm{c}_g) \right\|_F$, we set $\delta_{g} = \frac{8\pi^2}{\lambda^2}(\sum_{i, j, p, q=1}^L \sum_{k, k^{\prime}=1}^K|c_{ij}^{gkk'}||c_{qp}^{gkk'}| +2\sum_{i, j=1}^L \sum_{k=1}^K|e_{ij}^{gk} |)$. 	The minimum antenna spacing constraint in (\ref{obj_1c}) remains nonconvex. To address this, we apply a first-order Taylor approximation at $\bm{c}_{g}^{(i)}$. 
Combining  (\ref{loc_obj}),  the problem (\ref{opt_t}) can be expressed as \vspace{-5pt}
	\begin{align} 
		&{\max_{\bm{c}_g}} \ 
		 -\frac{\delta_{g}}{2} \bm{c}_g^T \bm{c}_g+\left(\nabla \mu\left(\bm{c}_g^{(i)}\right)+\delta_{g} \bm{c}_g^{(i)}\right)^T \bm{c}_g
		, \nonumber \\
		&{\rm{s.t.}}\   \frac{\left(\bm{c}_{g}^{(i)}-\bm{c}_{g'}\right)^{\rm t}\left(\bm{c}_{g}-\bm{c}_{g'}\right)}{\left\|\bm{c}_{g}^{(i)}-\bm{c}_{g'}\right\|}\geq D,  g\neq g',  (\ref{obj_1b}),\vspace{-10pt} 
	\end{align}
which is convex  and can be solved via the cvx tool.
\vspace{-10pt} 	
\section{Simulation Results}
  We analyze the effects of connectivity and movability  on the performance and results demonstrate the improvement brought by MAs over traditional FAs and BD-RIS with interconnections over conventional RIS. The distance between BS and BR-RIS is $d_{\rm BI} = 50 \ {\rm m}$, and $K=2$ UEs are randomly located within the circle centred at BD-RIS with radius  $d_{\rm IU} = 2  \ {\rm m}$. The geometric channel model is considered, where $L_{\mathrm{t}}=L_{\mathrm{r}}=L$ and  the PRM $\bm{\Sigma}$  is assumed to be diagonal \cite{Lzhu2025survey}, with  elements set as $[\bm{\Sigma}]_{1,1} \sim \mathcal{C} \mathcal{N}(0, \kappa /(\kappa+1))$ and $[\bm{\Sigma}]_{l, l} \sim \mathcal{C} \mathcal{N}\left(0,1 /\left((\kappa+1)\left(L-1\right)\right)\right)$ for $l=2,3, \ldots, L$. The ratio of the average power between LoS and NLoS paths  is set as $\kappa=1$. The path loss is defined  as $\eta(d)=\gamma_0 d^{-\alpha}$, where the signal attenuation at  unit distance is set to $\gamma_0 = -30 (dB)$   and the exponent is  $\alpha =2.2$. The noise power for all involved channels are $\sigma^2=-80 \rm dBm$. The azimuth and elevation angle of all channels are set as $\phi^{x}_{k,p}/ \theta^x_{k,p}\in [-\pi/2,\pi/2]$.
 The BD-RIS moving area $\mathcal{C}_{R}$ is a rectangle with length $l_1 = l_s l_{\rm FA}$ and width $l_2=4\lambda$, where $l_{\rm FA} = (M-1)\lambda/2$ represents the size of traditional FA and scaling factor $l_s$ decides the size of MA. The default  wavelength and scaling factor are $\lambda=0.01$ m and  $l_s=1.2$, respectively. The penalty and proximal parameters are $\rho = 0.5$ and $\xi = 0.1$.

\begin{figure}[H]
    \centering
    \begin{minipage}[H]{0.4\textwidth}
		\captionsetup{justification=raggedright, singlelinecheck=false}
        \includegraphics[width=\linewidth]{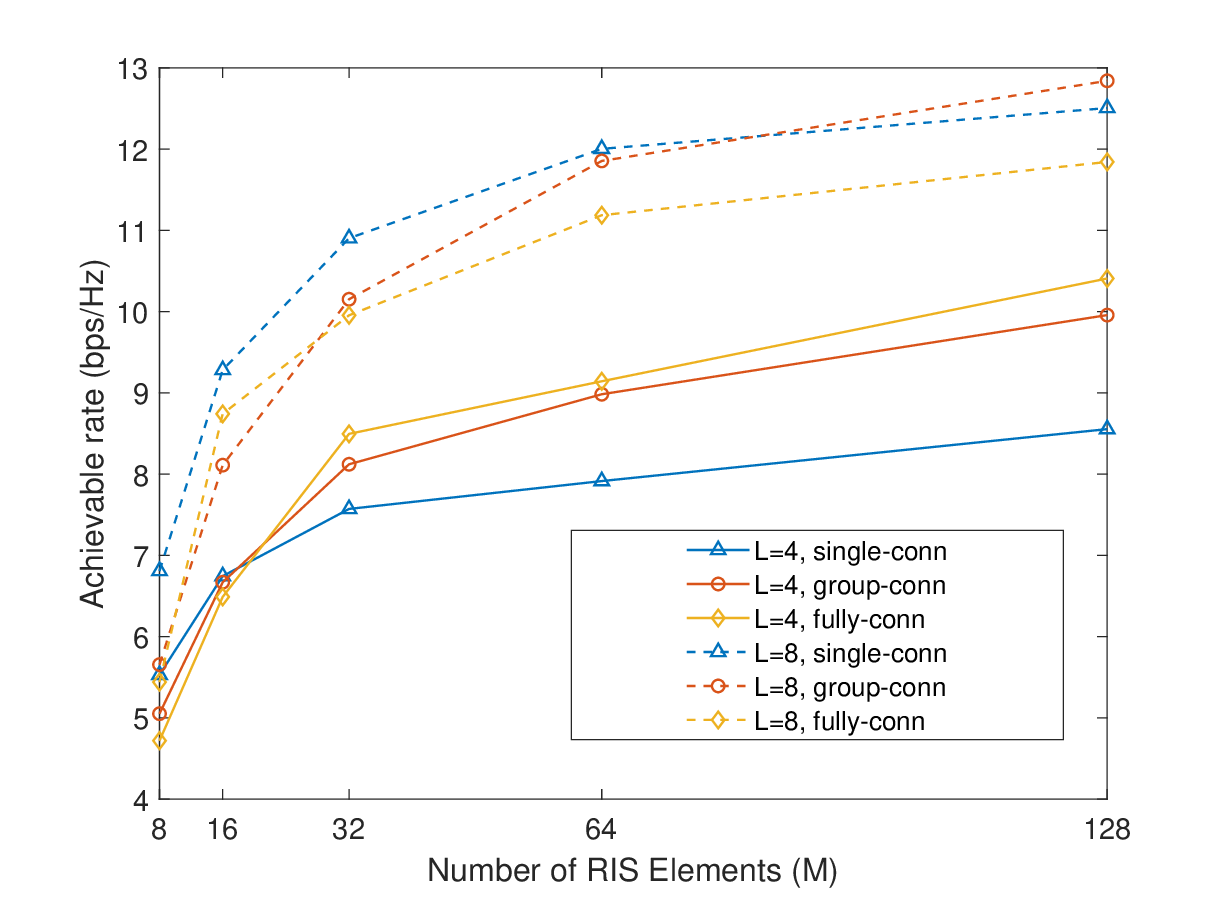}
        \caption{Achievable rate vs. number of  RIS elemnets  $M$, with $N_t = 4$.}
        \label{fig_l}
    \end{minipage}%
    \hfill
    \begin{minipage}[H]{0.4\textwidth}
		\captionsetup{justification=raggedright, singlelinecheck=false}
        \includegraphics[width=\linewidth]{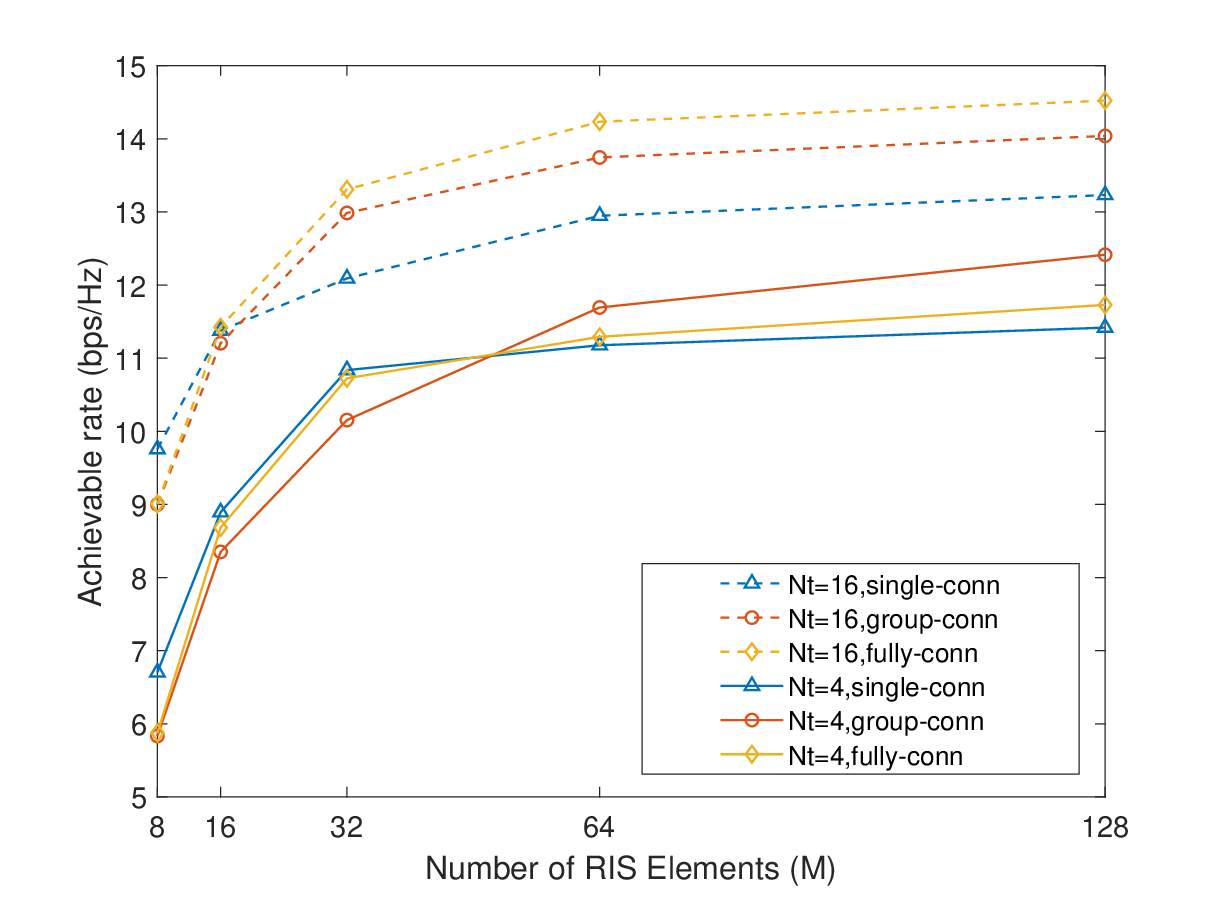}
        \caption{Achievable rate vs. number of RIS elemnets  $M$, with $L=6$.}
        \label{fig_Nt}
    \end{minipage}
\end{figure}

Figs. \ref{fig_l} and \ref{fig_Nt} explore the impact of the paths number $L$ and transmit antennas  $N_t$ on the performance of MA-assisted single-, group-, and fully-connected architectures for varying numbers of BD-RIS elements $M$. 
Key insights show a fundamental trade-off governed by  element number $M$. Movability yields greater performance gains when  $M$ is small. Conversely, a higher circuit connectivity offers larger gains as $M$ increases, exploiting the benefits of higher circuit degrees of freedom (DoF) at large scales.

Furthermore, as shown in Fig. 2, the highly-movable structure benefits more when the number of paths $L$ increases. Higher $L$ leads to rich scattering and severe spatial small-scale fading, making sufficient signal gain the performance bottleneck. The highly-movable structure has greater positional DoF, enabling it to place its elements at the instantaneous constructive interference peaks of the multipath field, thereby achieving superior spatial diversity gain.
\vspace{-23pt} 
\begin{figure}[H]
	\centering
        \includegraphics[width=0.4\textwidth]{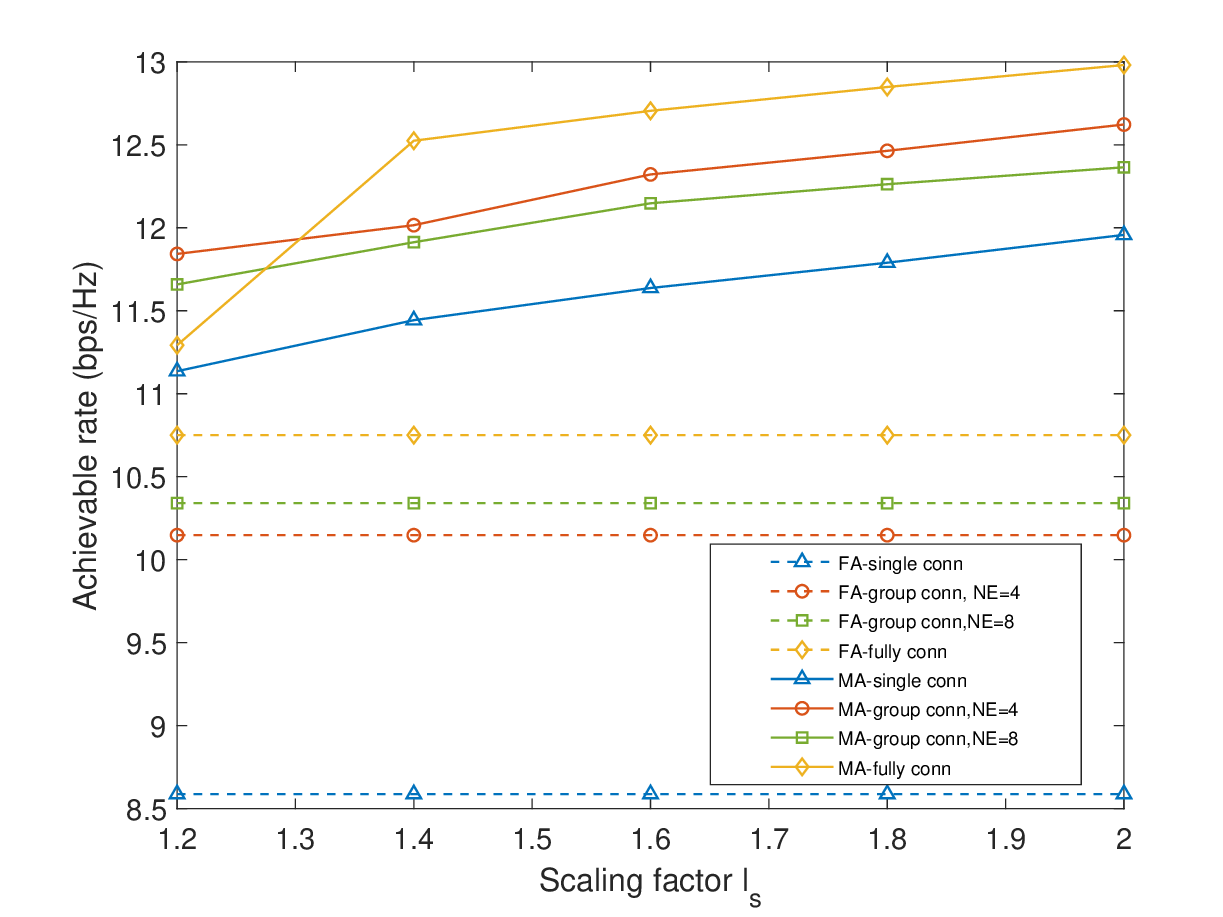}
     \caption{Achievable rate vs. scale factor $l_s$, with \!$M\!=\!64$, $L\!\!=\!6$, $N_t \!= \!4$.} \vspace{-15pt}
        \label{fig_area}
\end{figure} 
As shown in Fig.3, the highly-connected structure benefits more  when  $N_t$ increases. Higher $N_t$ yields higher beam resolution and correspondingly increased signal gain, shifting the performance bottleneck to efficient beam design. The highly-connected structure maximize its benefits by utilizing its superior circuit DoF for fine beamforming,
 leading to the highly-connected configuration performing  better.

Fig.~{\ref{fig_area}} compares the performance difference between MA-structure   and FA-structure under different moving ranges $l_s$ and group size $N_E$. It can be observed that even for a very small moving range, the MA-structure significantly outperforms the FA-structure. Besides,  since the group size simultaneously influences both connectivity and movability, there exits a  trade-off in selecting the optimal group size.
\section{Conclusion}
This work establishes a unified view of BD-RIS and movable antennas, showing that their integration enables a flexible balance between circuit connectivity and spatial movability. Through analytical modeling and simulation, we have revealed how the joint design of circuitry and geometry can influence the performance  of the MU-MISO systems. The observed trade-off between connectivity and movability not only deepens the understanding of BD-RIS-assisted networks but also provides a design reference for practical deployment.
\begin{appendices}
	\section{Calculation of $\mu(\bm{c_g})$}\label{app_a}
	Denote $R_1(\bm{c})=\sum_{k,k'=1}^{K}|{\psi}_k|^{2}|\bm{h}_{k}^{\rm H}(\bm{c})\bm{\Theta}\bm{H}(\bm{c})\bm{w}_{k'}|^2$ and $R_2(\bm{c})=\sum_{k=1}^{K}
		2\sqrt{1+\rho_k}
		\Re\{\bm{h}_{k}^{\rm H}(\bm{c})\bm{\Theta}\bm{H}(\bm{c})\bm{w}_{k}{\psi}_k^*\}$. Substituting    (\ref{channel2}) and (\ref{channel1})  into $R_1(\bm{c})$,  we have
			\vspace{-8pt} 
\begin{align}
	\!&R_1(\bm{c})
	\!\!=\!\!\!\!\!\sum_{k,k'=1}^{K}\!\!|\!
	\sum_{g = 1}^{N_G}\!{\rm Tr}(\!{\psi}_k^*\bm{1}^{\!\rm H}\!\bm{\Sigma}_k\!\bm{G}_{k}(\bm{c}_{g}\!)\bm{\Theta}_g\!
    \bm{F}^{\rm H}\!(\!\bm{c}_{g}\!)\!\bm{\Sigma}_{br}\bm{G}(\bm{b})\bm{w}_{k'})|^{2}
	\nonumber \\
	\!&\overset{(a)}{=}\!\!\!\!\!\sum_{k,k'=1}^{K}\!\!\!|\sum_{g = 1}^{N_G} \!\tilde{\bm{f}}^{\rm H}\!(\!\bm{c}_{g}\!)\tilde{\bm{C}}_{\!g\!,k\!,k'}\!\tilde{\bm{g}}_{k}\!(\!\bm{c}_{g}\!)\!|^2\!\!
	 \overset{(b)}{=}\!\!\!\!\!\sum_{k,k'=1}^{K}\!\!\!\!|\sum_{g = 1}^{N_G} \!\bar{\bm{f}}^{\rm H}\!(\!\bm{c}_{g}\!)\bar{\bm{C}}_{\!g\!,k\!,k'\!}\bar{\bm{g}}_{k}\!(\!\bm{c}_{g}\!)\!|^2
     \nonumber\\
	 \!	&\overset{(c)}{=}
		\!\!\!\!\!\sum_{k,k'=1}^{K}\!\!\!\!|\sum_{g = 1}^{N_G} \!{\bm{f}}^{\rm H}\!(\!\bm{c}_{g}\!){\bm{C}}_{\!g\!,k\!,k'\!}{\bm{g}}_{k}\!(\!\bm{c}_{g}\!)\!|^2\!\!
	\overset{(d)}{=}
		\!\!\!\!\!\sum_{k,k'=1}^{K}\!\!\!\!|\!{\bm{f}}^{\rm H}\!(\!\bm{c}_{g}\!){\bm{C}}_{\!g\!,k\!,k'\!}{\bm{g}}_{k}\!(\!\bm{c}_{g}\!)\!\!+\!a_{\!g\!,k\!,k'\!}^*|^2\!\!\!,\!
		\vspace{-15pt} 
\end{align}
where  eq. $(a)$ is based on 	${\rm Tr}(\bm{ABCD})=\operatorname{vec}(\bm{D}^{\rm T})^{\rm T}(\bm{C}^{\rm T} \otimes \bm{A}) \operatorname{vec}(\bm{B})$, thus $\tilde{\bm{f}}(\bm{c}_{g}) = {\rm vec}(\bm{F}(\bm{c}_{g}))\in \mathbb{C}^{L_r N_E\times 1}$, $\tilde{\bm{g}_k}(\bm{c}_{g})={\rm vec}(\bm{G}_k(\bm{c}_{g}))$, $\tilde{\bm{C}}_{g,k,k'} = \bm{\Theta}_g^{\rm T} \otimes(\bm{\Sigma}_{br}\bm{G}(\bm{b})\bm{w}_{k'}{\psi}_k^*
\bm{1}^{\rm H}\bm{\Sigma}_k)$. The eq. $(b)$ is established by the spatial relationship between the $m$-th FRV    $\bm{f}(\bm{t}_{g,m})$ and the reference FRV $\bm{f}(\bm{c}_{g})$, since  elements in the same group are uniformly spaced at half-wavelength intervals, thus we have 
\begin{equation} \label{a1}	
    \tilde{\bm{f}}(\bm{c}_{g}) = 
    \bar{\bm{A}}_g\bar{\bm{f}}(\bm{c}_{g}),
	\tilde{\bm{g}}(\bm{c}_{g}) = 
    \bar{\bm{B}}_g\bar{\bm{g}}(\bm{c}_{g}),
	\vspace{-3pt} 
\end{equation}	
where $
\bar{\mathbf{A}}_g = {\rm blkdiag}\{\mathbf{A}_{g,m}\}_{m=1}^{N_E}
$, 
with $\bm{A}_{g,m} = {\rm diag}\{e^{j\frac{2\pi}{\lambda}(\Delta x_{g,m}sin\phi_{p}^r cos \theta_{p}^r +\Delta y_{g,m}sin \theta_{p}^r)}\}_{p=1}^{L_{r}}\in\mathbb{C}^{L_{r} \times L_{r}}$ and $\bar{\bm{f}}(\bm{c}_{g}) = \mathbf{1}_{N_E} \otimes \bm{f}(\bm{c}_{g})$. The same definition for $\bar{\bm{B}}_g$ and $\bar{\bm{g}}(\bm{c}_{g})$, thus $\bar{\bm{C}}_g  = \bar{\bm{A}}_g^{\rm H}\tilde{\bm{C}}_g\bar{\bm{B}}_g$. The eq. $(c)$ results from the block matrix multiplication, thus ${\bm{C}}_g = \sum_{i=1}^{N_E}\sum_{j=1}^{N_E}\bar{\bm{C}}_{g,i,j}$, where $\bar{\bm{C}}_{g,i,j}\in \mathbb{C}^{L_r \times L_t}$ is the $(i,j)$-th submatrix of $\bar{\bm{C}}_g$. The eq. $(d)$ reorganize the formula and $a_g =\sum_{i = 1,i\neq g}^{N_G} ({\bm{f}}^{\rm H}(\bm{c}_{i}){\bm{C}}_i{\bm{g}}(\bm{c}_{i}))^{\rm H}$.  
Following the same reasoning above, we can obtain 
\begin{equation}\label{a2}
	 R_2(\bm{c})=\sum_{g = 1}^{N_G} \sum_{k=1}^{K}2\Re\{ 
		{\bm{f}}^{\rm H}(\bm{c}_{g}){\bm{D}}_{g,k}{\bm{g}_{k}}(\bm{c}_{g})\},
			\vspace{-3pt} 
\end{equation}
where ${\bm{D}}_{g,k}$ is calculated in the same way as $\bm{C}_g$, based on $\tilde{\bm{D}}_{g,k} = \bm{\Theta}_g^{\rm T} \otimes(\sqrt{1+\rho_k}\bm{\Sigma}_{br}\bm{G}(\bm{b})\bm{w}_k{\psi}_k^*
		\bm{1}^{\rm H}\bm{\Sigma}_k)$. 
	Combining (\ref{a1}) and (\ref{a2}) and ignoring the constant term, the objective function can be rewritten as (\ref{u_g}) with  ${\bm{E}}_{g,k} =-\sum_{k'=1}^{K}a_{g,k,k'}\bm{C}_{g,k,k'}+\bm{D}_{g,k}$.	
\end{appendices}

\begin{thebibliography}{99}
	\bibitem{QWUTCOM2021} Q. Wu, S. Zhang, B. Zheng, C. You and R. Zhang,`` Intelligent Reflecting Surface-Aided Wireless Communications: A Tutorial," {\em IEEE Trans. Commun.}, vol. 69, no. 5, pp. 3313--3351, May 2021,
	\bibitem{szhang2020jsac}S. Zhang and R. Zhang,``Capacity Characterization for Intelligent Reflecting Surface Aided MIMO Communication," {\em IEEE J.
Select. Areas Commun.}, vol. 38, no. 8, pp. 1823--1838, Aug. 2020

	

	\bibitem{ZWangtwc2024}Z. Wang, X. Hu, C. Liu and M. Peng, ``RIS-Enabled Multi-Target Sensing: Performance Analysis and Space-Time Beamforming Design,"  {\em IEEE Trans. Wireless Commun.}, vol. 23, no. 10, pp. 13889--13903, Oct. 2024


	\bibitem{Lzhu2025survey}L. Zhu et al., ``A Tutorial on Movable Antennas for Wireless Networks," {\em IEEE Commun. Surveys $\&$ Tuts.}, 2025. 

\bibitem{Ghulett2024} G. Hu et al., “Intelligent reflecting surface-aided wireless communication
with movable elements,” {\em IEEE Wireless Commun. Lett.} , vol. 13, no. 4,
pp. 1173--1177, 2024.

  \bibitem{Yzhang2024arxiv} Y. Zhang, I. Dey, and N. Marchetti, “RIS-aided wireless communication
with movable elements geometry impact on performance,”  {\em arXiv preprint
arXiv} :2405.00141, 2024.

 \bibitem{Hlijsac2023}H. Li, S. Shen and B. Clerckx, ``Beyond Diagonal Reconfigurable Intelligent Surfaces: A Multi-Sector Mode Enabling Highly Directional Full-Space Wireless Coverage," {\em IEEE J.
Select. Areas Commun.}, vol. 41, no. 8, pp. 2446--2460, Aug. 2023,
	


	\bibitem{SSHENTWC2022}S. Shen, B. Clerckx, and R. Murch,`` Modeling and architecture design
	of reconfigurable intelligent surfaces using scattering parameter network
	analysis,''{\em IEEE Trans. Wireless Commun.}, vol. 21, no. 2, pp. 1229--1243,
	Feb. 2022.
	
 \bibitem{Hliarxiv2025}H. Li, M. Nerini, S. Shen, B.Clerckx,  ``A Tutorial on Beyond-Diagonal Reconfigurable Intelligent Surfaces: Modeling, Architectures, System Design and Optimization, and Applications.'' {\em arXiv preprint
arXiv}: 2505.16504, 2025.

\bibitem{Nerini2023letter}M. Nerini and B. Clerckx, ``Pareto Frontier for the Performance-Complexity Trade-Off in Beyond Diagonal Reconfigurable Intelligent Surfaces,"  {\em IEEE Commun. Lett.}, vol. 27, no. 10, pp. 2842--2846, Oct. 2023
	\bibitem{zheyuarxiv2024}Z. Wu, B. Clerckx, ``Optimization of Beyond Diagonal RIS: A Universal Framework Applicable to Arbitrary Architectures,'' {\em arXiv preprint
arXiv}:  2412.15965, 2024.


	\bibitem{ShenTSP2018} K. Shen and W. Yu, ``Fractional programming for communication systems—Part I: Power control and beamforming,'' {\em IEEE Trans. Signal Process.}, vol. 66, no. 10, pp. 2616--2630, May. 2018.



		\bibitem{Stephenadmm} S. Boyd, N. Parikh, E. Chu, B. Peleato, and J. Eckstein, ``Distributed optimization and statistical learning via the alternating direction method of multipliers,'' {\em Foundations Trends Mach. Learn.}, vol. 3, no. 1, pp. 1--122, 2011.


\end{thebibliography}
\end{document}